\documentclass[a4paper]{panl}
\usepackage{float}
\usepackage{cite}
\usepackage{wrapfig}
\usepackage{graphicx}
\usepackage{amssymb}
\usepackage{amsfonts}
\usepackage{amsmath}
\usepackage{longtable}
\usepackage{rotating}
\usepackage{lscape}
\usepackage{epsfig}
\usepackage{multirow}
\usepackage{subcaption} 
\usepackage{lineno}
\usepackage{placeins}
\originalTeX

\newcommand{\mean}[1]{\left\langle #1 \right\rangle}
\begin{document}

\issuearea{Physics of Elementary Particles and Atomic Nuclei. Experiment}

\title{The comparison of methods for anisotropic flow measurements with the MPD Experiment at NICA}

\maketitle

\authors{P.~Parfenov $^{a}$, A.~Taranenko $^{a}$ \footnote{E-mail: AVTaranenko@mephi.ru},
  D.~Idrisov $^{a}$, V.~B ~Luong $^{a}$, N.~Geraksiev $^{b,c}$}
\authors{A.~Demanov $^{a}$, A.~Povarov $^{a}$,
	V. Kireyeu$^{c}$, A.~Truttse $^{a}$, E. Volodihin$^{a}$}
\authors{for the MPD Collaboration}
\from{$^{a}$\,National Research Nuclear University MEPhI, Moscow, Russia\\
$^{b}$\,Plovdiv University “Paisii Hilendarski”, Plovdiv, Bulgaria\\
$^{c}$\,VBLHEP, Joint Institute for Nuclear Research, Dubna, Russia}

\begin{abstract}

The anisotropic collective flow is one of the key observables to study the properties of dense
matter created in heavy-ion collisions. The performance
of Multi-Purpose Detector (MPD) at NICA collider for directed and elliptic
flow measurements is studied with Monte-Carlo simulations of heavy-ion collisions
at energies $\sqrt{s_{NN}}$ = 4 - 11 GeV.
\end{abstract}


\section{Introduction}

\begin{figure}[htb]
	\centering
	\includegraphics[width=0.45\textwidth] {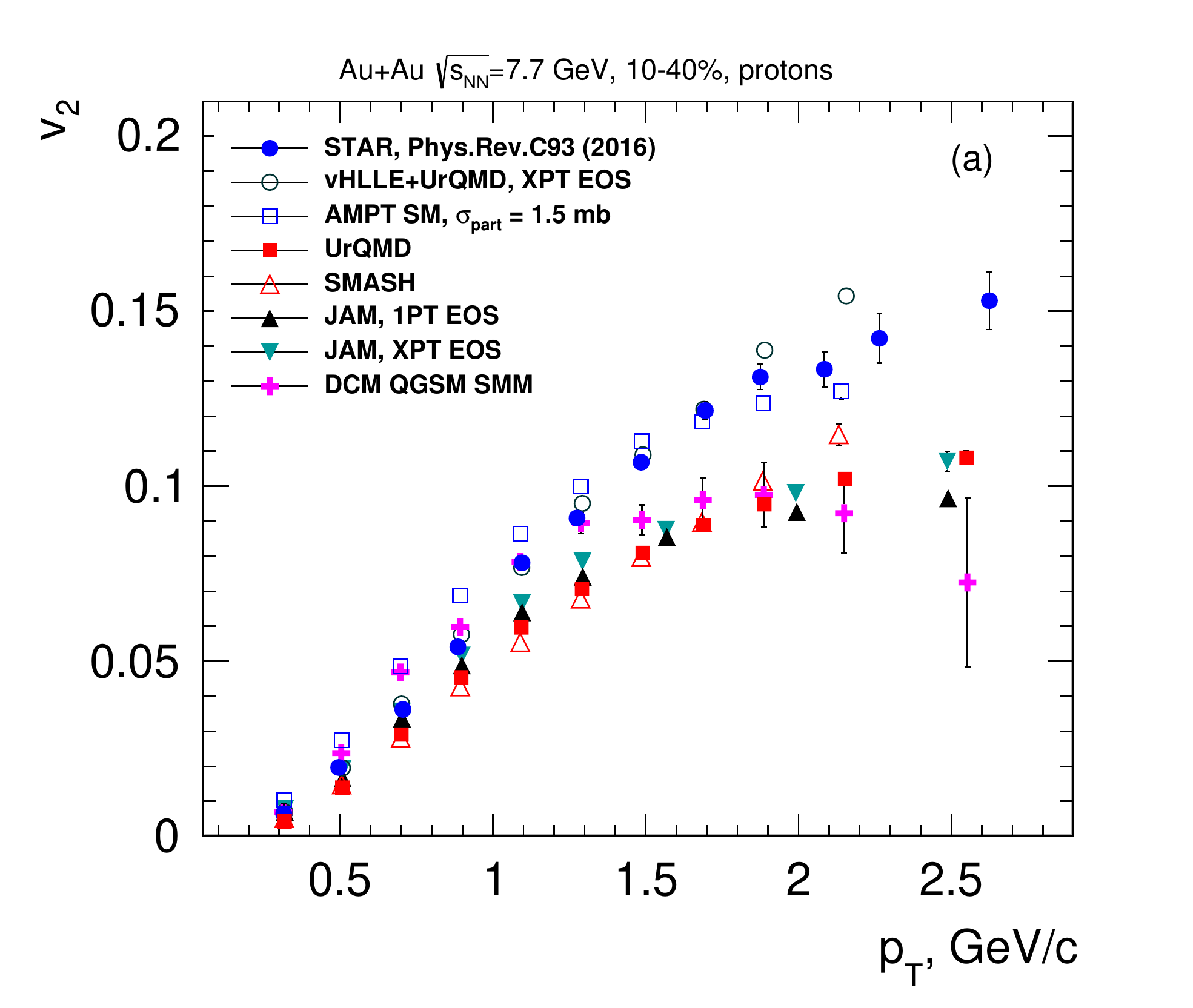}
	\includegraphics[width=0.45\textwidth] {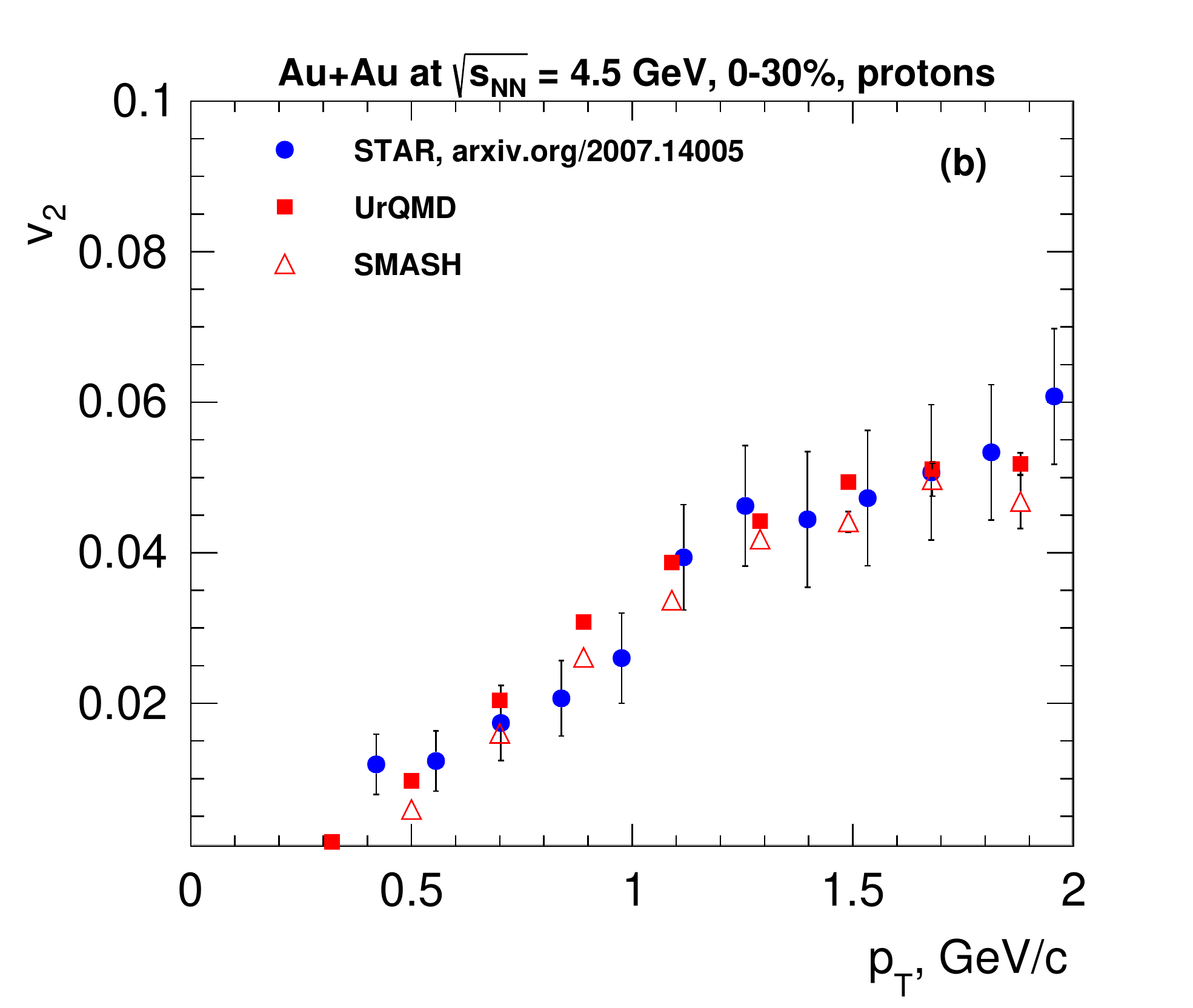}
	\caption{$p_T$ dependence of elliptic flow of protons from 10-40\% midcentral Au+Au collisions at
          $\sqrt{s_{NN}}=7.7$ GeV (left) and 0-30\% central Au+Au collisions at
          $\sqrt{s_{NN}}=4.5$ GeV (right). Blue closed circles represent the published data from STAR
          experiment \cite{starv2pid1,starv2pid2}
          and other symbols
        the results from event plane analysis of generated  events from UrQMD, SMASH, JAM , AMPT
        and hybrid vHLLE+UrQMD models. }
	\label{fig:v2protSTAR}
\end{figure}

The Multi-Purpose Detector (MPD) at NICA collider has a substantial discovery potential concerning
the exploration of the QCD phase diagram in the region of high net-baryon densities and moderate
temperatures \cite{nica,mpd}.  The anisotropic collective flow, as manifested 
by the anisotropic emission of particles in the plane transverse to the beam 
direction, is one of the important observable sensitive to the
properties of the strongly interacting 
matter: the equation of state (EOS), the specific shear  and bulk
viscosity \cite{flow2}. It can be quantified by the 
Fourier coefficients $v_n$ in the expansion of the particles azimuthal distribution 
as: $dN/d\phi \propto 1 + \sum_{n=1} 2 v_{n} \cos 
(n(\phi-\Psi_{n}))$, where $n$ is the order of 
the harmonic, $\phi$ is the azimuthal angle of particles of a given type, 
and $\Psi_n$ is the azimuthal angle of the $n$th-order event plane. In this work,  we briefly review
the available experimental results for the collision energy dependence of 
directed ($v_1$) and elliptic ($v_2$) flow and
discuss the anticipated performance of MPD detctor for flow measurements at NICA energies.
The directed flow ($v_1$) can probe the very early stages
of the collision as it is 
generated  during the passage time of the two colliding nuclei
$t_{pass}=2R/(\gamma_s \beta_s)$, where $R$ is the radius of the nucleus
at rest, $\beta_s$ is the spectator velocity in c.m. and $\gamma_s$ the 
corresponding Lorentz factor, respectively.  Both hydrodynamic  and transport model
calculations indicate that the $v_1$ signal  of  baryons,
is sensitive to the equation of state \cite{v1review} and predict a minimum in  $dv_1/dy$  for the 
first order phase transition  between hadronic matter
and sQGP \cite{v1review}.  The recent results from the Beam Energy Scan (BES-I) program  of
STAR experiment at RHIC
show a minimum at
$\sqrt{s_{NN}}$ = 10-20 GeV for $dv_1/dy$ for protons and $\Lambda$ hyperons
from midcentral Au+Au collisions \cite{v1bes}. Further progress in the area
of model calculations and high-statistics differential measurements of $v_1$
is needed to find the reason for such non-monotonic behavior.\\
The published data from STAR experiment shows that $v_2(p_T)$ for charged hadrons
changes relatively little as a function of beam energy in the range $\sqrt{s_{\small{NN}}}$ = 11.5 - 62.4 GeV
\cite{starv2pid1} and this may result from the interplay of the hydrodynamic and hadronic
transport phase \cite{Karpenko:2015}.  In the energy range $\sqrt{s_{\small{NN}}}$ = 11 - 2 GeV,
the passage time $t_{pass}$ increases from 2 fm/c to 16 fm/c and  the shadowing
effects by the spectator matter start to  play an important role for the generation of
elliptic flow \cite{jam2bes}. The left part of Fig.~\ref{fig:v2protSTAR} shows $v_2(p_T)$ of 
protons from 10-40\% midcentral Au+Au collisions at $\sqrt{s_{NN}}=7.7$ GeV.  Blue closed circles
represent the published data from STAR experiment \cite{starv2pid1}
and other symbols the results from event plane analysis of generated  events from the current
state of the art models of heavy-ion collisions:
UrQMD~\cite{Bleicher:1999xi,Bass:1998ca}, SMASH \cite{smash}, JAM \cite{jam2bes}, DCM-QGSM-SMM \cite{dcm},
AMPT \cite{ampt} and hybrid vHLLE+UrQMD \cite{Karpenko:2015}.
We found  that, hybrid models with QGP formation:
viscous hydro + hadronic cascade  vHLLE+UrQMD model \cite{Karpenko:2015}
or string melting version of AMPT \cite{ampt}
provide a relatively good description of $v_2(p_T)$ of protons  in Au+Au collisions
at $\sqrt{s_{NN}}=7.7$ GeV and above.  Pure hadronic transport models:
UrQMD, SMASH, JAM and DCM-QGSM-SMM: generally underpredicts the $v_2$ values. However, the situation
is different for Au+Au collisions at $\sqrt{s_{NN}}=4.5$ GeV, see right part of Fig.~\ref{fig:v2protSTAR}.
Here, the pure hadronic transport system (as described by the UrQMD and SMASH models)
appear to explain the measurements for $v_2(p_T)$ of protons from the STAR experiment \cite{starv2pid2}.
The high-statistics differential measurements of $v_n$
anticipated from the MPD experiment at NICA expected to
provide  valuable information about this parton-hadron transient energy domain.

\vspace{-0.5pc}

\section{The MPD detector system at NICA}
\begin{figure}[htb]
	\centering
	\includegraphics[width=0.35\textwidth] {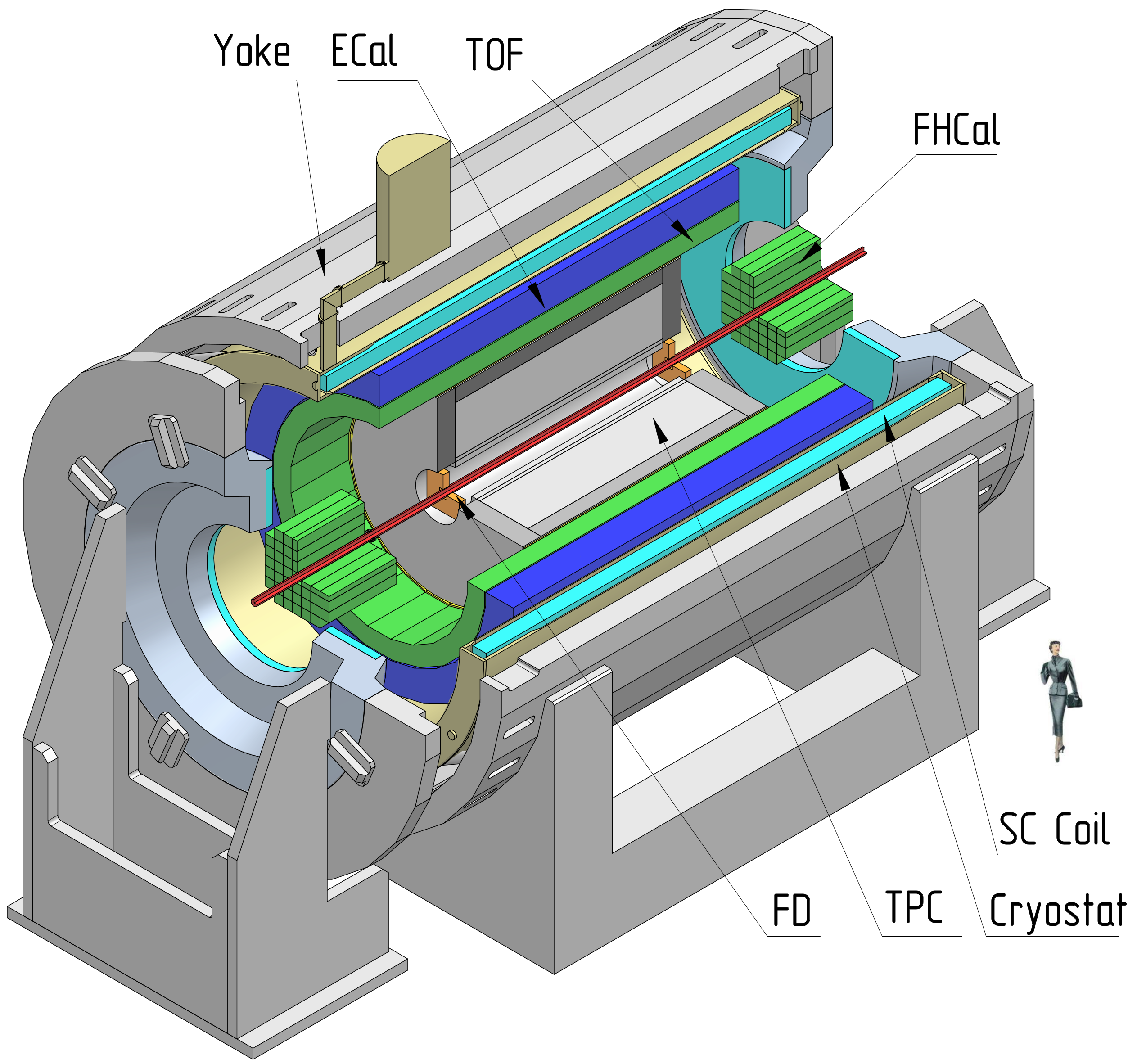}
	\includegraphics[width=0.4\textwidth] {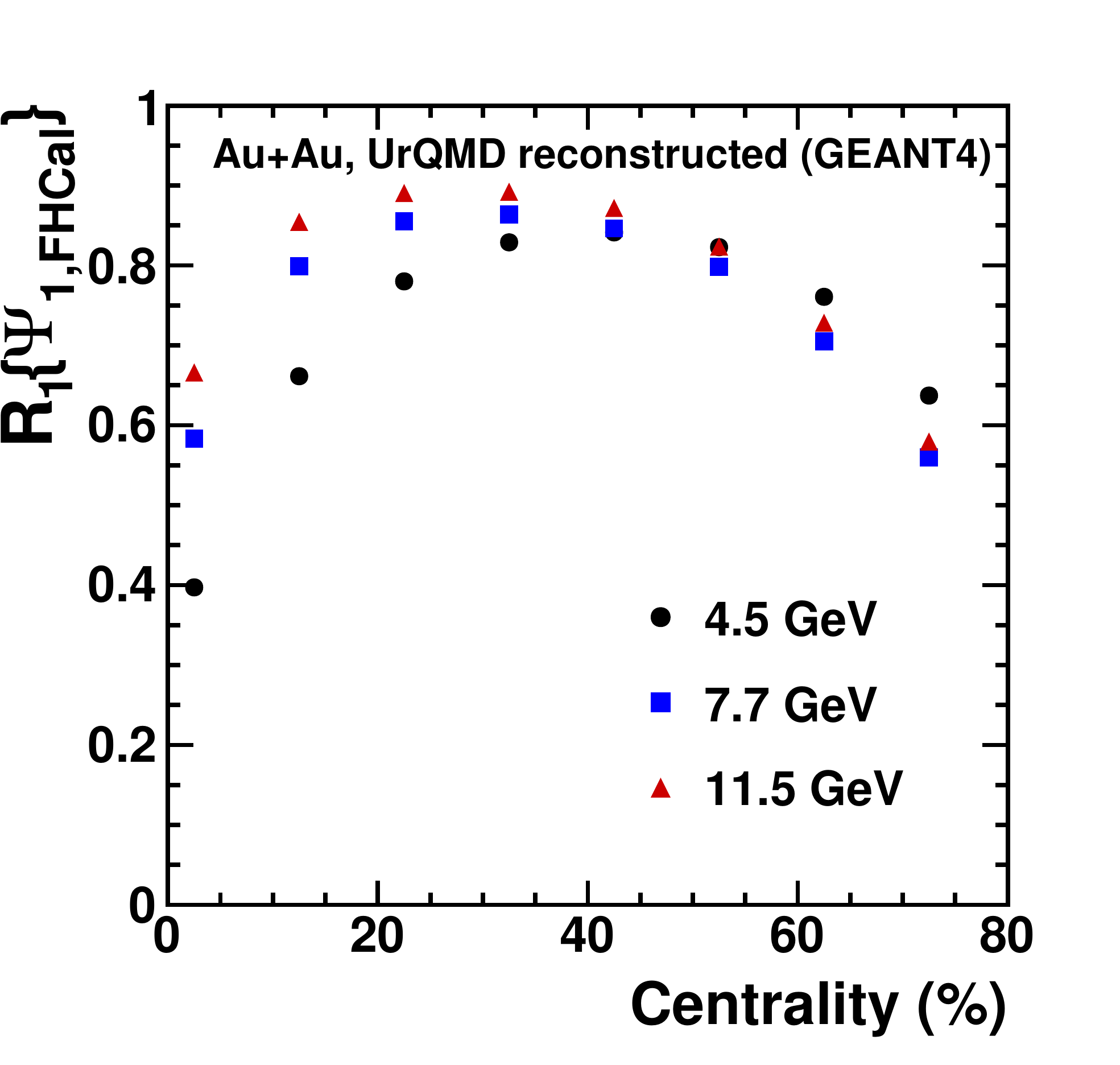}
	\caption{(left) The schematic view of the MPD detector in Stage 1. (right)
		Centrality dependence of event plane resolution factors
		$\rm R_1(\Psi_{1,\textrm{FHCal}})$ for $v_1$ measurements.}
	\label{fig:res}
\end{figure}

\vspace{-0.5pc}
The MPD detector system (Fig.~\ref{fig:res}, left) consists of a barrel part and two endcaps located inside the
magnetic field. Time Projection Chamber (TPC) will be the central MPD tracking detector \cite{mpd}. TPC
will provide 3D tracking of charged particles, as well as the
measurement of specific ionization energy loss dE/dx for particle identification for $|\eta|<$ 1.2.
The TPC will be surrounded by a cylindrical barrel of the Time-of-Flight (TOF) detector with a timing resolution of
the order of 50 ps.  The combined system TPC+TOF will allow the efficient charged
pion/kaon separation up to 1.5 GeV/c and protons/meson separation up to 2.5 GeV/c.
The Forward Hadronic Calorimeter
(FHCal), placed at 2$<|\eta|<$ 5, will be used for centrality determination as well
for the reconstruction of event plane from the directed flow of particles.
\vspace{-0.5pc}

In this work, we use  cascade version of
UrQMD model~\cite{Bleicher:1999xi,Bass:1998ca}
to simulate the heavy-ion collisions at NICA energies.
In total, the sample of 120 M of minimum bias Au+Au events at $\sqrt{s_{NN}}=7.7$ and 11.5 GeV was
used for directed and elliptic flow performance study using different methods of analysis. We used
term ``true'' $v_n$ data for these results. At the next step, a sample of
10-25 M UrQMD minimum bias events, depending on the analysis,  was
used as an input for the full chain of the realistic simulations of the MPD
detector subsystems' based on the GEANT4 platform and reconstruction algorithms build in the MPDROOT.
We named these $v_n$ results as the ``reco'' $v_n$ data.
The main workflow for the analysis of identified charged hadrons with the
reconstructed data is similar to  the previous work~\cite{Parfenov:2019pxf}.
For $K_{s}^{0}$ and $\Lambda$ particles analysis
the  secondary vertexes are reconstructed using a Kalman filtering algorithm based on the
MpdParticle paradigm by combining identified decay products with  set of topological cuts
to optimize the signal \cite{zinchenko}.

\section{Methods for anisotropic flow measurements in MPD}
\vspace{-0.5pc}

\begin{figure}[htb]
	\centering
	\includegraphics[width=0.33\textwidth] {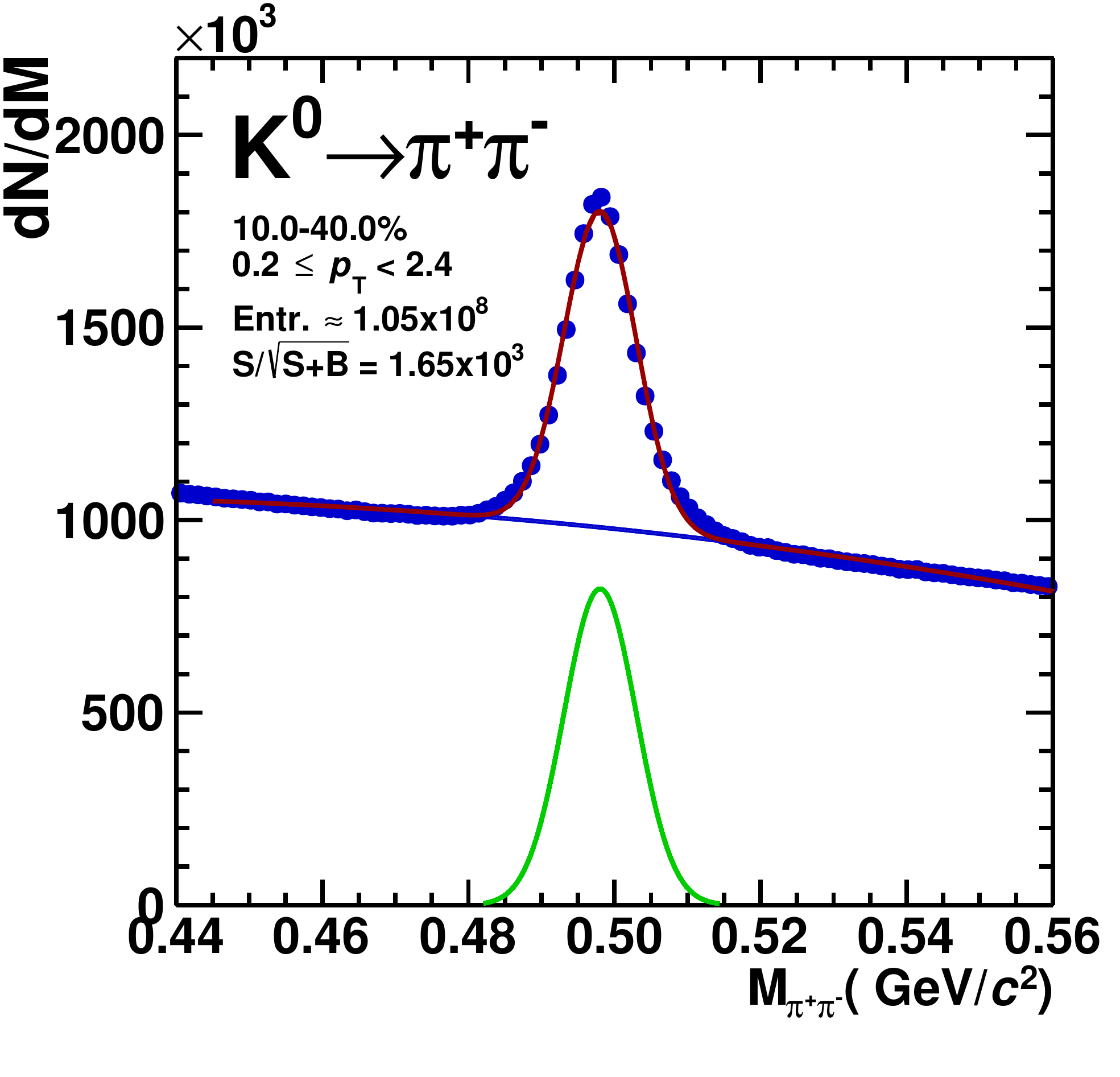}
	\vspace{-0.5pc}
	\includegraphics[width=0.33\textwidth] {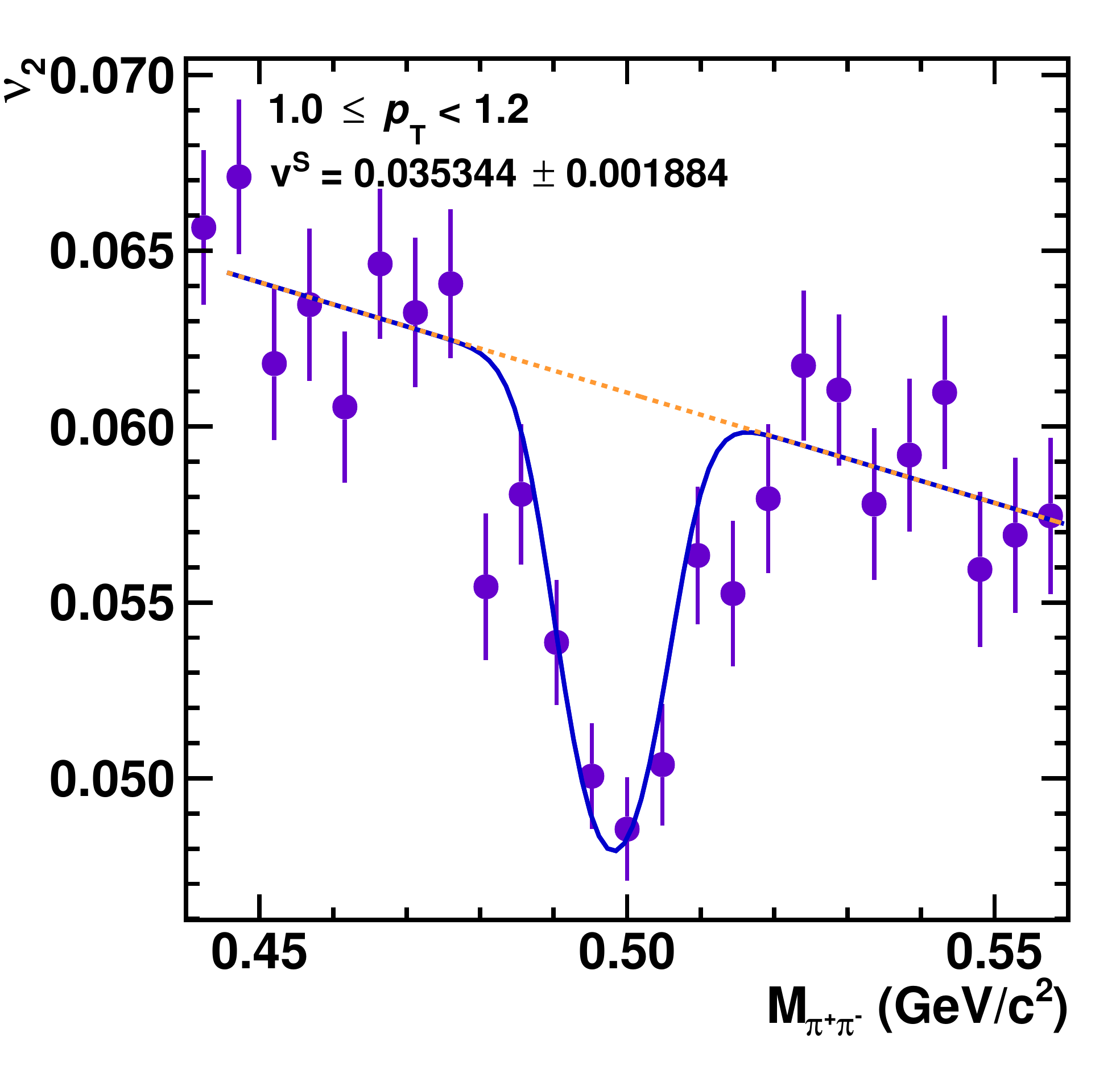} \\
    \includegraphics[width=0.33\textwidth] {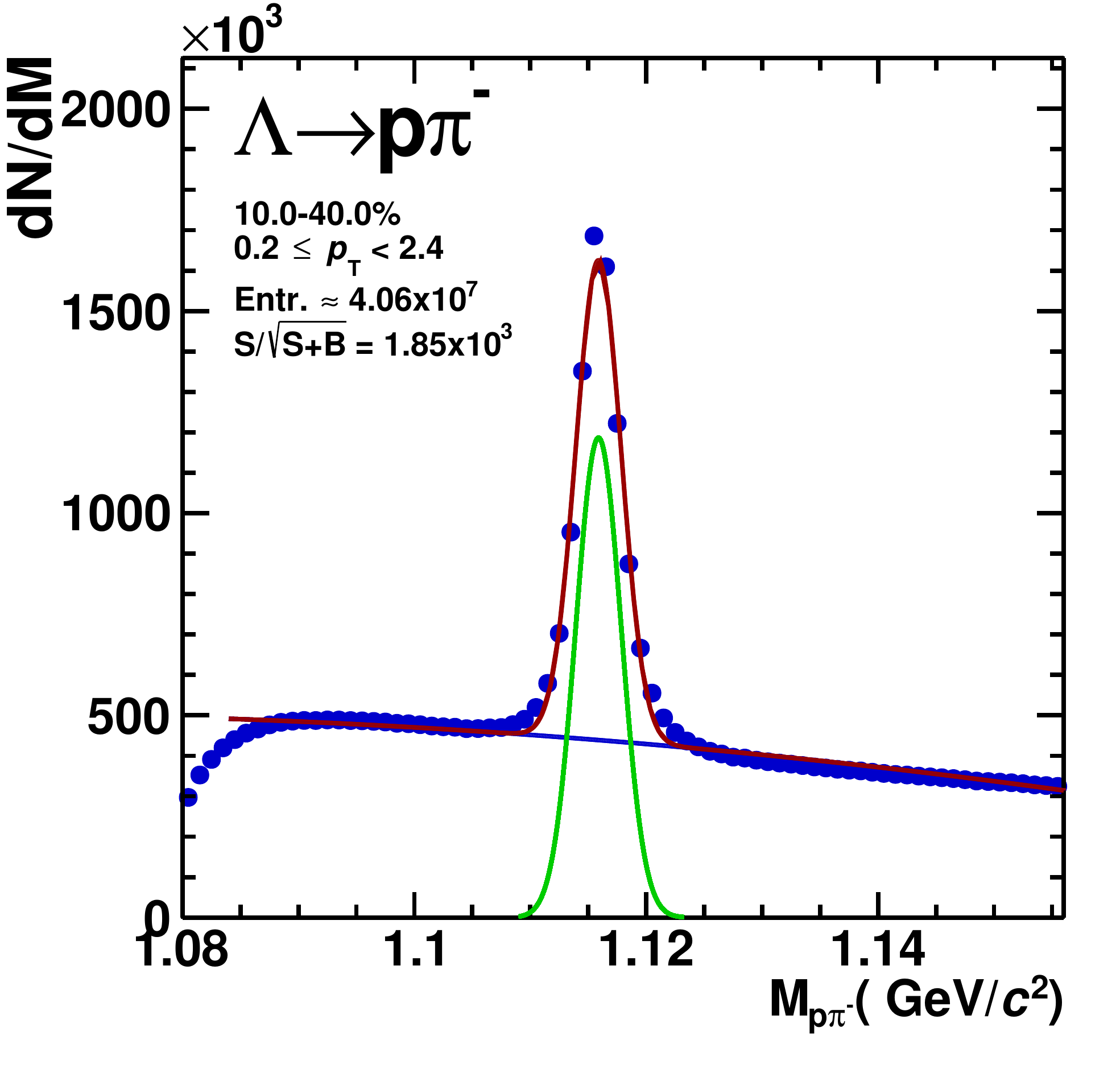}
	\includegraphics[width=0.33\textwidth] {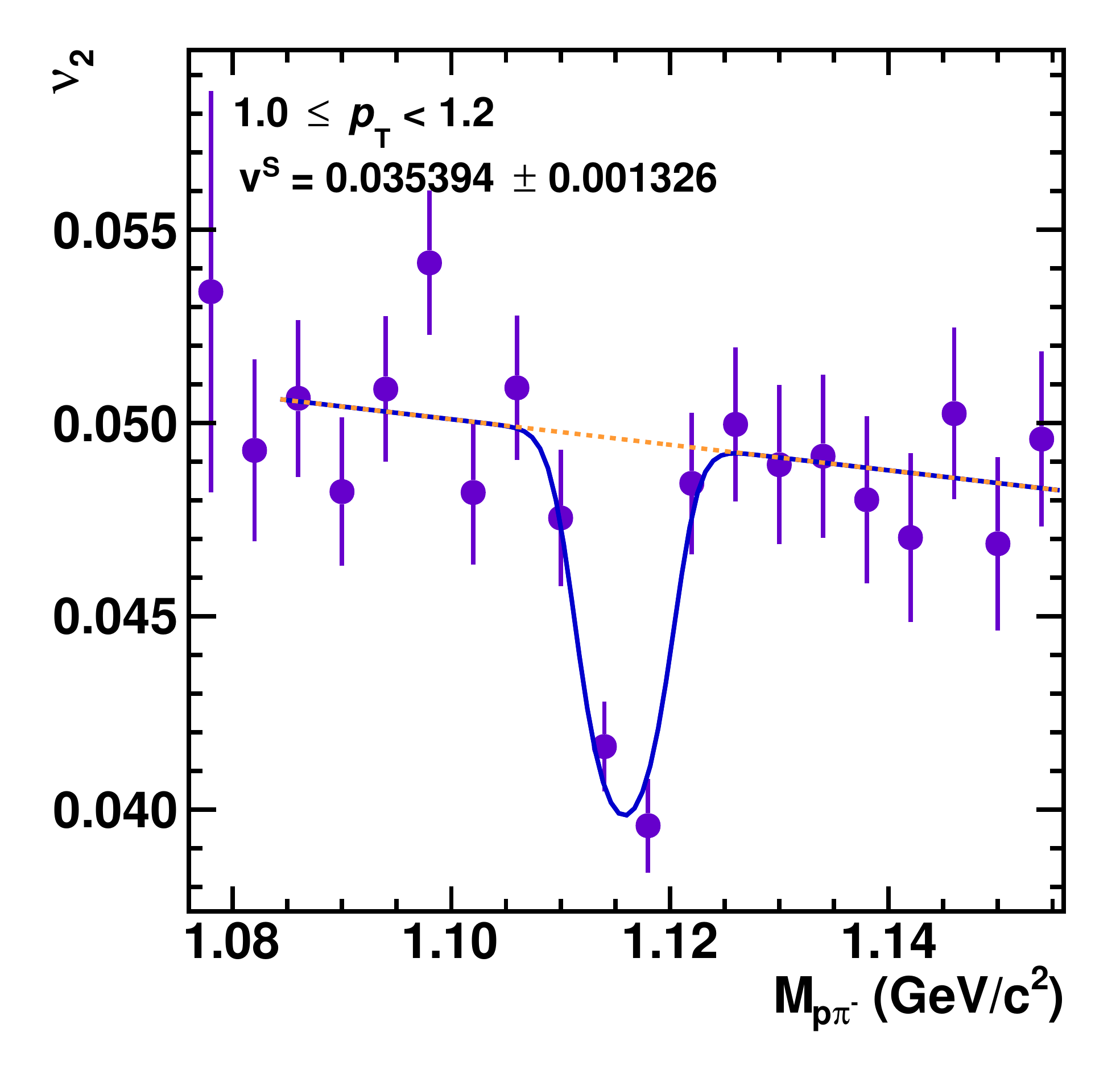}

	\caption{left: Invariant mass distributions for  $K_{S}^{0}$ (upper part) and $\Lambda$ (lower part)
          particles
          from 10-40\% midcentral Au+Au collisions at $\sqrt{s_{NN}}=11$~GeV. right: the demonstration
          of invariant-mass fit method for extraction of $v_{2}^{S}$ signal for $K_{S}^{0}$ and $\Lambda$ particles.}
	\label{fig:v2ksfit}
\end{figure}

In this section, we discuss how the event plane, scalar product  and direct cumulant methods can be used for the measurements of
anisotropic  flow of the produced particles with MPD detector system at NICA.\\
The event plane method  correlates  azimuthal angle $\phi$ of each particle
with the azimuthal angle $\Psi_n$ of event plane determined from the anisotropic
flow itself \cite{vol2008}.  The event flow vector ($Q_n$) and the azimuthal angle of event plane $\Psi_n$ can be defined 
for each harmonic, $n$, of the Fourier expansion by:
\begin{eqnarray}
	Q_{n,x} = \sum \limits_{i} \omega_{i} \cos (n\varphi_i),\ Q_{n,y} = \sum \limits_{i} \omega_{i} \sin (n\varphi_i), \Psi_n = \frac{1}{n} \tan^{-1} \left( \frac{Q_{n,y}}{Q_{n,x}} \right),
\end{eqnarray}
where the sum runs over all particles $i$ used in the event plane calculation, and $\varphi_i$  and  $\omega_i$ are the
laboratory azimuthal angle and the weight for the particle $i$. In this work we use two estimators for 
event plane:  $\Psi_{1,\textrm{FHCal}}$ determined from the
directed flow (n=1) of particles detected in the FHCal (2$<|\eta|<$ 5) and   $\Psi_{2,\textrm{TPC}}$
determined from the elliptic flow (n=2) of produced particles detected in the TPC ($|\eta|<$ 1.5).
The reconstructed $\Psi_{1,\textrm{FHCal}}$ can be used for the measuremented of directed (n=1) and elliptic  (n=2) flow
$v_{n}^{\Psi_{1,\text{FHCal}}}$
of the produced
particles, detected in TPC.  $\Psi_{2,\textrm{TPC}}$ allows to get an independent estimate of elliptic flow $v_{2}^{\Psi_{2,\text{TPC}}}$.

\begin{equation}
v_{2}^{\Psi_{2,\text{TPC}}}=\frac{\langle\cos(2(\phi_{i} - \Psi_{2,\text{TPC}}))\rangle}{R_{2}(\Psi_{2,\text{TPC}})}, \:
v_{n}^{\Psi_{1,\text{FHCal}}}=\frac{\langle\cos(n(\phi_{i} - \Psi_{1,\text{FHCal}}))\rangle}{R_{n}(\Psi_{1,\text{FHCal}})}, 
\end{equation}

where $\rm R_2(\Psi_{2,\textrm{TPC}})$ and $\rm R_n(\Psi_{1,\textrm{FHCal}})$ represent the resolution of the event planes.
\vspace{-0.5pc}
\begin{figure}[htb]
	\centering
	\includegraphics[width=0.40\textwidth] {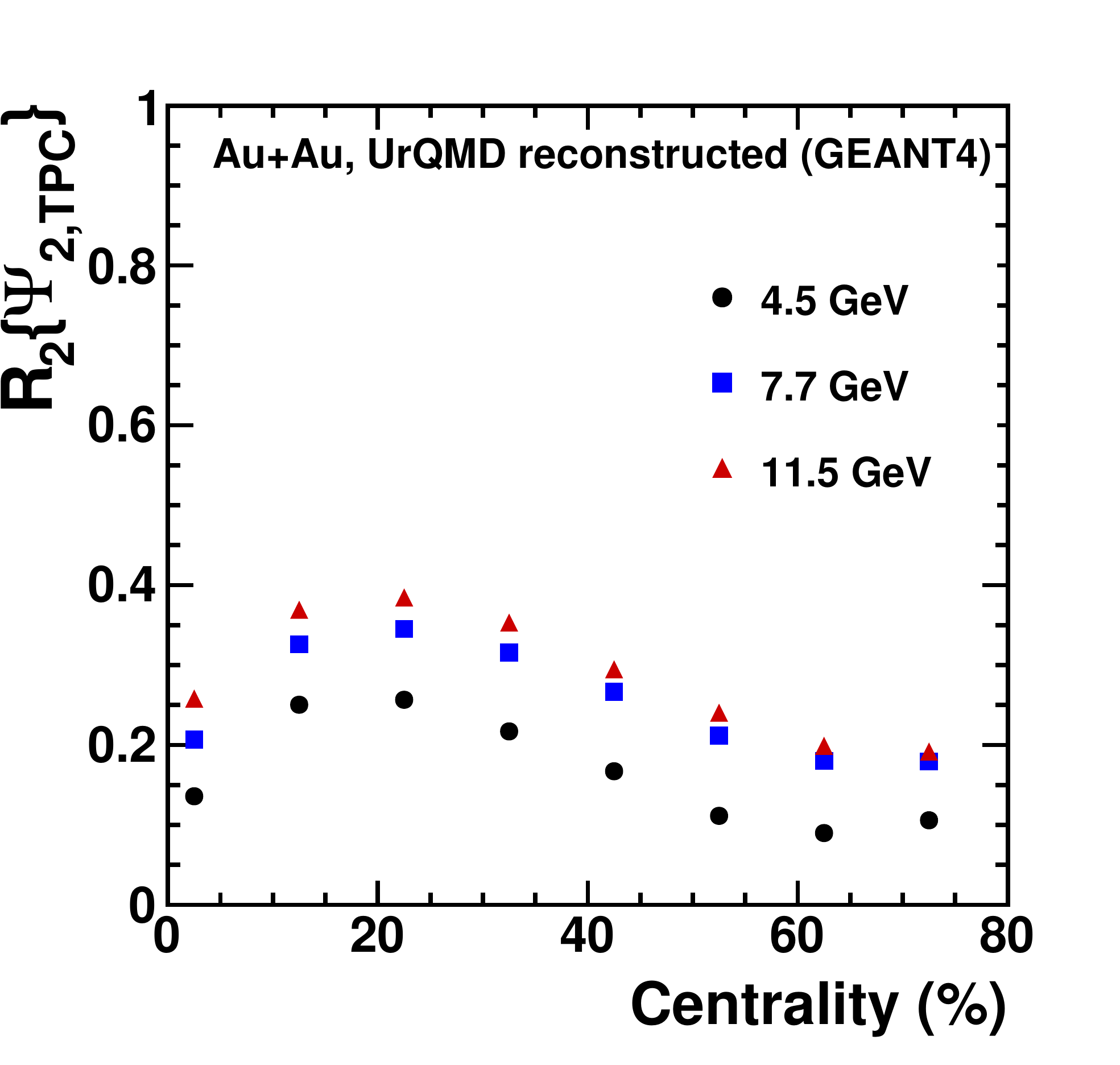}
	\includegraphics[width=0.40\textwidth] {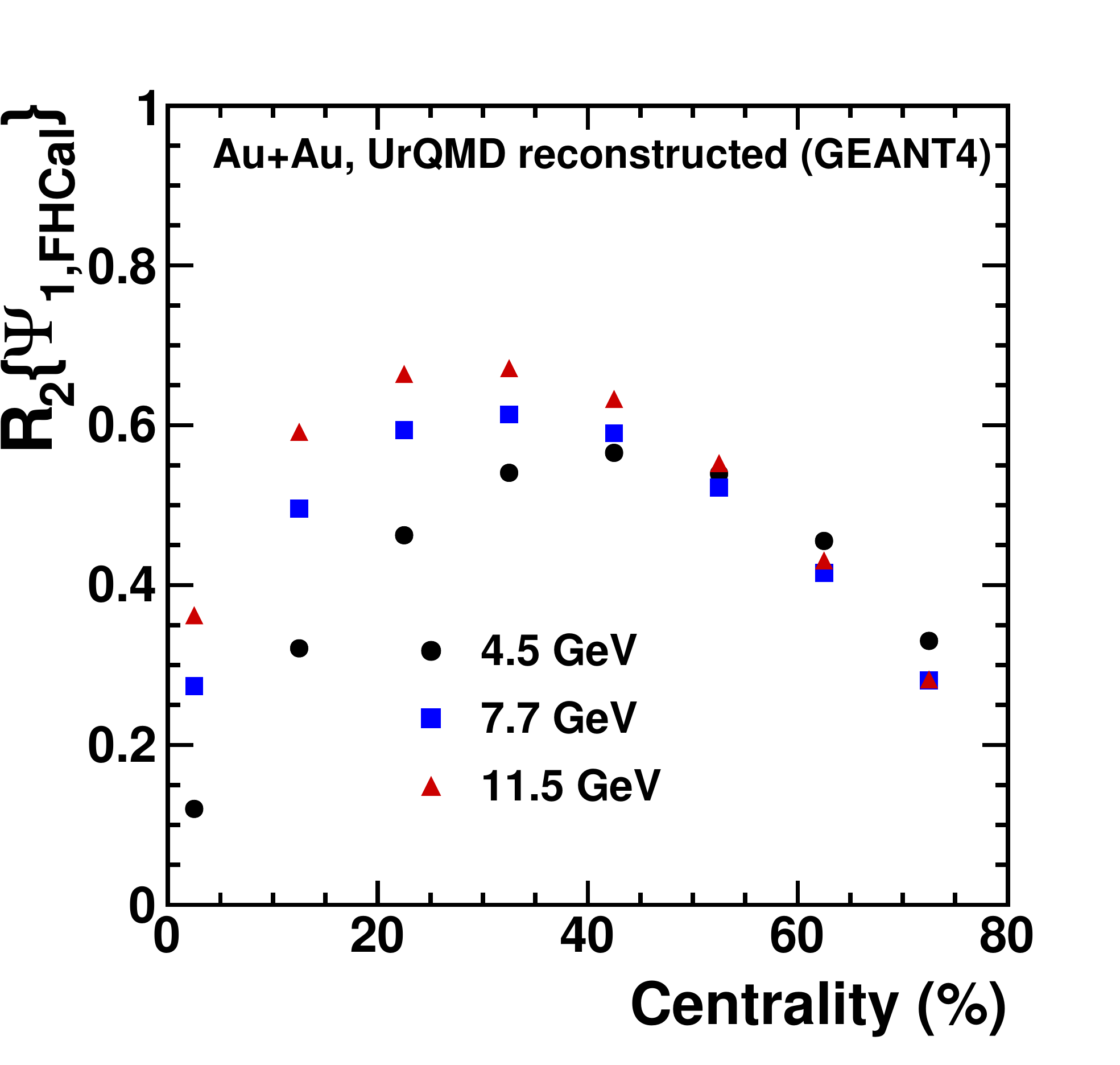}
	\vspace{-0.5pc}
	\caption{Centrality dependence of event plane resolution factors $\rm R_2(\Psi_{2,\textrm{TPC}})$ (left)
		and   $\rm R_2(\Psi_{1,\textrm{FHCal}})$ (right) for Au+Au collisions at $\sqrt{s_{NN}}$ = 4.5, 7.7 and 11.5 GeV.}
	\label{fig:resol}
\end{figure}
%
The right part of Fig.~\ref{fig:resol}  shows the centrality dependence of $\rm R_1(\Psi_{1,\textrm{FHCal}})$  for the
directed flow measurements with respect to $\Psi_{1,\textrm{FHCal}}$ plane
for Au+Au collisions at $\sqrt{s_{NN}}$ = 4.5, 7.7 and 11 GeV. The results are
based on the analysis of the fully reconstructed  UrQMD events. The centrality dependence of 
$\rm R_2(\Psi_{2,\textrm{TPC}})$ and
$\rm R_2(\Psi_{1,\textrm{FHCal}})$ for elliptic flow measurements in presented in Fig.~\ref{fig:resol}.\\
In the scalar product method (SP) for
differential flow $v_n(p_T)$ measurements one uses 
the magnitude of the flow vector ($Q_n$) as a weight \cite{vol2008}:
\begin{eqnarray}
	\label{eq:flow_SP}
	v_n^\textrm{SP}\{Q_{n,\textrm{TPC}}\}(p_T)
        = \left\langle u_{n,i}(p_T) Q_n^* \right\rangle /2\sqrt{\left\langle Q_n^a Q_n^{b*} \right\rangle },
\end{eqnarray}
where $u_{n,i}$ is the unit vector of the $i^{th}$ particle (which is not included in $Q_n$ vector) and $a$
and $b$ are two subevents. If $Q_n$ vector is replaced by its unit vector, the scalar product method reduces to
event plane method. In this work we present the $v_2$ results obtained by SP method. \\
For $K_{s}^{0}$ and $\Lambda$ particles, the $v_{n}^{SB}$ of selected sample contains
both  $v_{n}^{S}$ of the signal and the $v_{n}^{B}$ of 
background \cite{flowfit}.  Therefore, the $v_{n}^{SB}$ is  measured as a function of invariant mass ($M_{inv}$) and
$p_T$:
\begin{eqnarray}
v_{n}^{SB}(M_{inv},p_{T}) =v_{n}^{S}(p_T)\frac{N^{S}(M_{inv},p_T)}{N^{SB}(M_{inv},p_T)} + 
v_{n}^{B}(M_{inv},p_T)\frac{N^{B}(M_{inv},p_T)}{N^{SB}(M_{inv},p_T)}
\end{eqnarray}
where $N^{S}(M_{inv},p_T)$, $N^{B}(M_{inv},p_T)$ and $N^{SB}(M_{inv},p_T)$
are signal, background and total yields obtained for each $p_T$ interval
from fits to the $K_{s}^{0}$ and $\Lambda$ invariant mass distributions, see left panels
of Fig.~\ref{fig:v2ksfit}. As an example the right panels of Fig.~\ref{fig:v2ksfit} 
illustrate the procedure of extraction of $v_{2}^{S}$. Values for $v_{2}^{S}$ signal for
$K_{S}^{0}$ and $\Lambda$ particles  were extracted via direct fit to the $v_{2}^{SB}(M_{inv})$
for each $p_T$ selection by Eq.4, see right panels of Fig.~\ref{fig:v2ksfit}. That is, the background
$v_{2}^{B}(M_{inv})$ was parametrized as a linear
function of $M_{inv}$  and $v_{2}^{S}$ was taken as a fit parameter.

In the Q-cumulant method the two- and four- particle cumulants (for each harmonic $n$)
can be calculated directly from a $Q_n$ vector, constructed
using particles from the TPC acceptance $|\eta|<$ 1.5,
$Q_n \equiv \sum_{i}^{M} exp \left( i n \varphi_i \right)$~\cite{Bilandzic:2010jr}:
\begin{gather}
	\left\langle 2 \right\rangle_n 
	= (\left| Q_n \right|^2 - M) /M(M-1),
	\\
	\left\langle 4 \right\rangle_n 
	= \frac{\left| Q_n \right|^4 + \left| Q_{2n} \right|^2 - 2 \Re [Q_{2n}Q_n^*Q_n^*] -4(M-2)\left| Q_n \right|^2 - 2M(M-3) }{M(M-1)(M-2)(M-3)}.
\end{gather}
M denotes the multiplicity in each
event used in the analysis. The elliptic flow ($n=2$) can be defined via the Q-cumulant method as follows:
\begin{eqnarray}
	v_2\{2\} = \sqrt{\left\langle \left\langle 2 \right\rangle \right\rangle},\ v_2\{4\} = \sqrt[4]{2\left\langle \left\langle 2 \right\rangle \right\rangle^2 - \left\langle \left\langle 4 \right\rangle \right\rangle},
\end{eqnarray}
where the double brackets denote weighted  average over all events. Equations for the $p_T$-differential elliptic  flow can be
found in~\cite{Bilandzic:2010jr}. \\

Different methods of analysis can be affected by nonflow and flow fluctuations in different ways. The 
nonflow effects are mainly due to few particle correlations, not associated with the
reaction plane: Bose-Einstein correlations, resonance decays,
momentum conservation. In this work we discuss  the comparison of different methods for elliptic flow only.
The estimates of $v_2$ 
based on multi-particle cumulants have the advantage of significant
reduction of contribution $\delta_2$ from
nonflow effects: $\left\langle 2 \right\rangle_2 = v_2^2 + \delta_2,\ \left\langle 4 \right\rangle_2 = v_2^4 + 4 v_2^2\delta_2 + 2\delta_2^2$. In order to suppress nonflow effects in $v_2$ results from two particle
correlation methods one can use rapidity gaps between correlated particles. For
$v_2\{\Psi_{2,\textrm{TPC}}\}$, $v_2^\textrm{SP}\{Q_{2,\textrm{TPC}}\}$, $v_2\{2\}$ we use the
$\eta$-gap of $\Delta\eta >$ 0.1 between the two sub-events. The  $v_2\{\Psi_{1,\textrm{FHCal}}\}$ results
are expected to be less affected by
nonflow due to larger  $\eta$-gap between particles in TPC and FHCal:
$\Delta\eta >$ 0.5. \\
Anisotropic  flow can fluctuate event to event. We define the elliptic flow fluctuations by
$\sigma_{v2}^2 = \mean{v_2^2}-\mean{v_2}^2$. Here, the resulting flow signal,
averaged over all events is denoted as $\left\langle v_2 \right\rangle$.
In the case of the Q-cumulants ($v_2\{2\}$ and $v_2\{4\}$),  for a Gaussian model of fluctuations  and in the limit 
$\sigma_{v2}\ll \mean{v_2}$  one can write~\cite{Voloshin:2007pc,vol2008}:
\begin{eqnarray}
	\label{eq:fluctuation_cumulants}
	v_2\{2\} = \left\langle v_2 \right\rangle + 0.5\cdot\sigma_{v2}^2/\left\langle v_2 \right\rangle,\ v_2\{4\} = \left\langle v_2 \right\rangle - 0.5\cdot\sigma_{v2}^2/\left\langle v_2 \right\rangle.
\end{eqnarray}
\begin{figure}[htb]
	\centering
	\vspace{-0.25in}
	\includegraphics[width=0.96\textwidth] {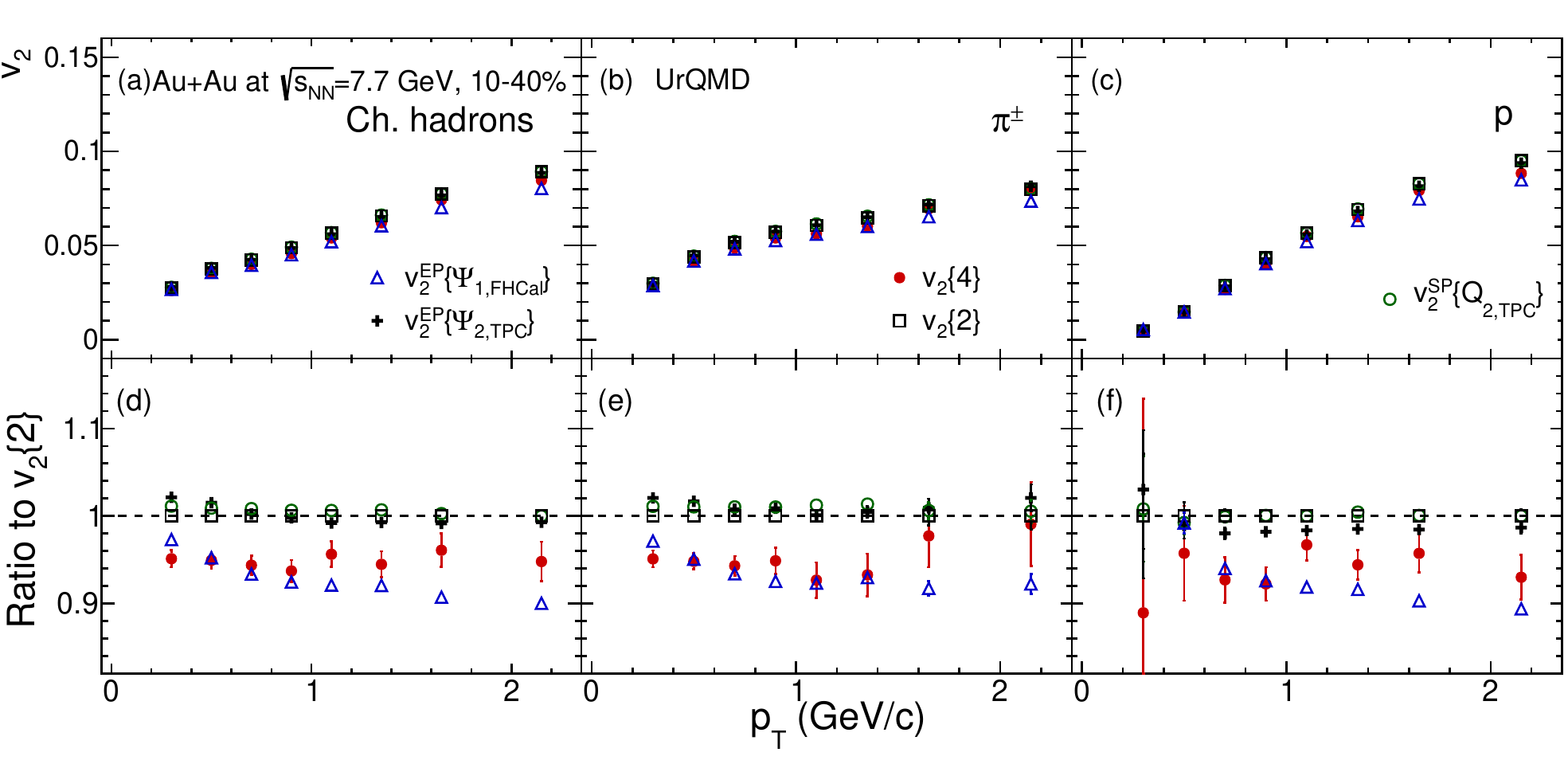}
	\vspace{-0.5pc}
	\caption{ $p_T$-dependence of $v_2$ of inclusive charged hadrons (a), pions (b) and protons (c) from
		10-40\% midcentral Au+Au collisions at $\sqrt{s_{NN}}=7.7$~GeV obtained using the  event
		plane ($v_2\{\Psi_{1,\textrm{FHCal}}\}$, $v_2\{\Psi_{2,\textrm{TPC}}\}$), scalar product
		$v_2^{SP}\{Q_{2,\textrm{TPC}}\}$) and Q-cumulant
		($v_2\{2\}$, $v_2\{4\}$) methods.
		Lower row shows the ratio $v_2$(method)/$v_2\{2\}$.}
	\label{fig:methods_compare}
\end{figure}
One of the important sources of $v_2$ flow fluctuations are participant eccentricity fluctuations in the initial geometry
of the overlapping region of two colliding  nuclei. Therefore, the $v_2\{\Psi_{1,\textrm{FHCal}}\}$ values
are expected to  be smaller than $v_2\{\Psi_{2,\textrm{TPC}}\}$ measured with respect to
the participant plane $\Psi_{2,\textrm{TPC}}$~\cite{Voloshin:2007pc,vol2008}:
\begin{eqnarray}
	\label{eq:fluctuation_EP}
	v_2\{\Psi_{1,\textrm{FHCal}}\} \simeq \left\langle v_2 \right\rangle,\ v_2\{\Psi_{2,\textrm{TPC}}\} \simeq \left\langle v_2 \right\rangle + 0.5\cdot\sigma_{v2}^2/\left\langle v_2 \right\rangle.
\end{eqnarray}

Figure~\ref{fig:methods_compare} shows the $p_T$ dependence of $v_2$ of inclusive charged hadrons,
charged pions and protons  from 10-40\% midcentral Au+Au collisions at $\sqrt{s_{NN}}=7.7$~GeV.
Different symbols correspond to the the $v_2$ results obtained by event
plane ($v_2\{\Psi_{1,\textrm{FHCal}}\}$, $v_2\{\Psi_{2,\textrm{TPC}}\}$), scalar product
$v_2^{SP}\{Q_{2,\textrm{TPC}}\}$) and Q-cumulant ($v_2\{2\}$, $v_2\{4\}$) methods of
analysis of events from UrQMD model. The ratios of $v_2$ signal to the $v_2\{2\}$ are
shown on the bottom panels and show good agreement between $v_2$ results obtained
by $v_2\{\Psi_{2,\textrm{TPC}}\}$, $v_2^{SP}\{Q_{2,\textrm{TPC}}\}$ and  $v_2\{2\}$ methods.
Both $v_2\{4\}$ and $v_2\{\Psi_{1,\textrm{FHCal}}\}$ methods give a smaller $v_2$ signal as one expect
from elliptic flow fluctuations and nonflow effects.

\section{Results}

The event plane ($v_n\{\Psi_{1,\textrm{FHCal}}\}$, $v_n\{\Psi_{2,\textrm{TPC}}\}$) and Q-cumulant ($v_n\{2\}$, $v_n\{4\}$) methods 
were implemented in the MPDROOT framework.
\begin{figure}[htb]
	\centering
	\includegraphics[width=0.96\textwidth] {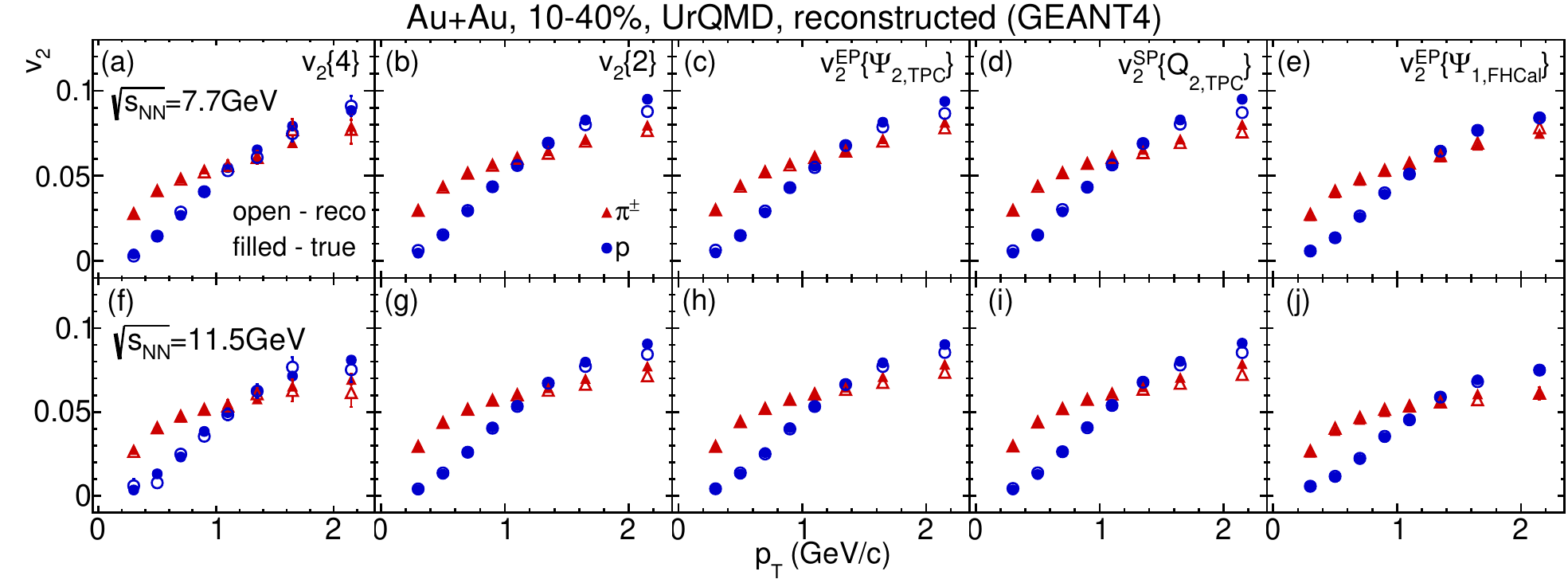}
	\vspace{-0.1pc}
	\caption{ Comparison of $v_2(p_T)$ for charged pions and protons from
		10-40\% midcentral  Au+Au collisions at $\sqrt{s_{NN}}=7.7$~GeV (upper panels) and
		$\sqrt{s_{NN}}=11.5$~GeV (lower panels)
		obtained  by Q-cumulant, event plane, scalar product  methods of analysis of 
		fully reconstructed  ("reco") and generated  UrQMD events ("true").}
	\label{fig:performance}
\end{figure}
Figure~\ref{fig:performance} shows the $p_T$ dependence
of $v_2$ of charged pions and protons  from 10-40\% midcentral  Au+Au collisions at $\sqrt{s_{NN}}=7.7$~GeV (upper panels) and 
$\sqrt{s_{NN}}=11.5$~GeV (lower panels).  The perfect agreement between
$v_2$ results from the analysis of fully reconstructed ("reco") and generated ("true") UrQMD events is observed.

\begin{figure}[htb]
	\centering
	\vspace{-0.14in}
	\includegraphics[width=0.45\textwidth] {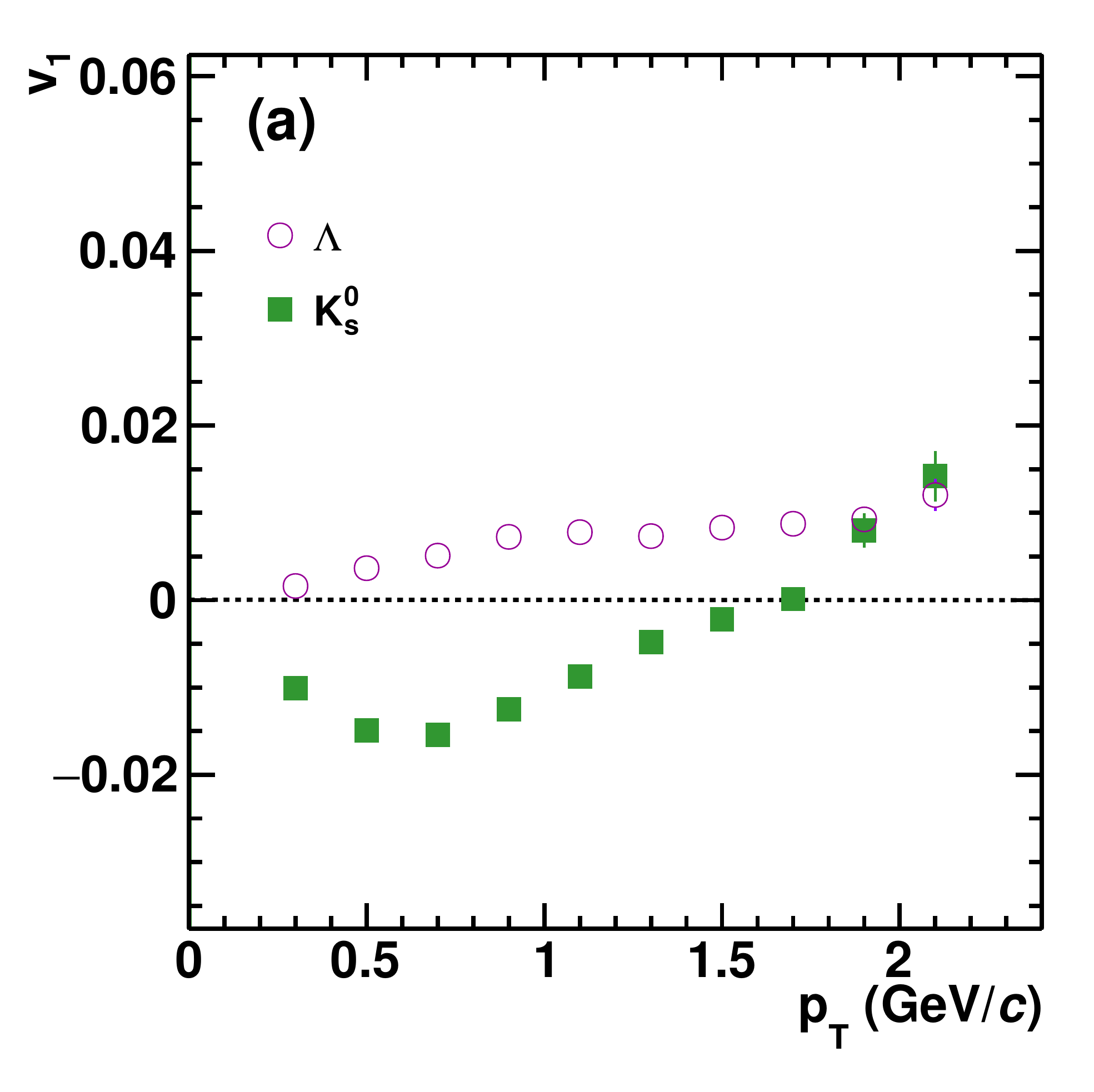}
	\includegraphics[width=0.45\textwidth] {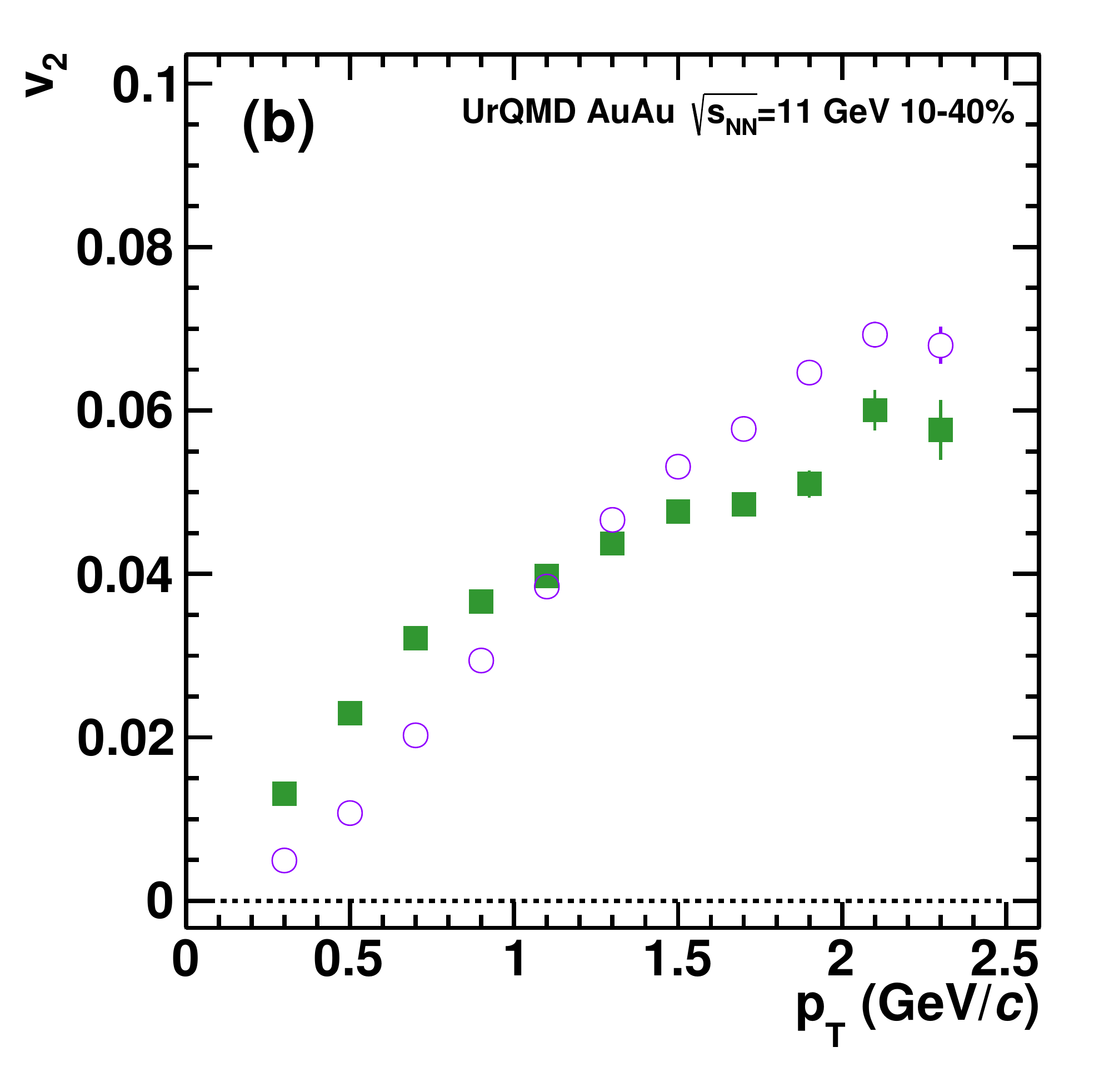}
	\caption{ $p_T$-dependence of directed (a) and elliptic (b) flow of $K_S^0$ and $\Lambda$ particles
          from 10-40\% midcentral Au+Au collisions at $\sqrt{s_{NN}}=11.$~GeV.
          The results were obtained by the invariant-mass fit method of
          the fully reconstructed UrQMD events.}
	\label{fig:v1v2kslambda}
\end{figure}

Figure~\ref{fig:v1v2kslambda} illustrates the  MPD detector system's performance
for the $p_T$ differential directed and elliptic flow measurements of $K_{S}^{0}$  and $\Lambda$ 
particles from 10-40\% midcentral Au+Au collisions at $\sqrt{s_{NN}}=11$~GeV.
The results were obtained from the event plane 
analysis of 25M minimumbias fully reconstructed UrQMD events using the
invariant-mass fit method, illustrated in the Fig.~\ref{fig:v2ksfit}.
\begin{figure}[htb]
	\centering
	\vspace{-0.14in}
	\includegraphics[width=0.9\textwidth,clip]{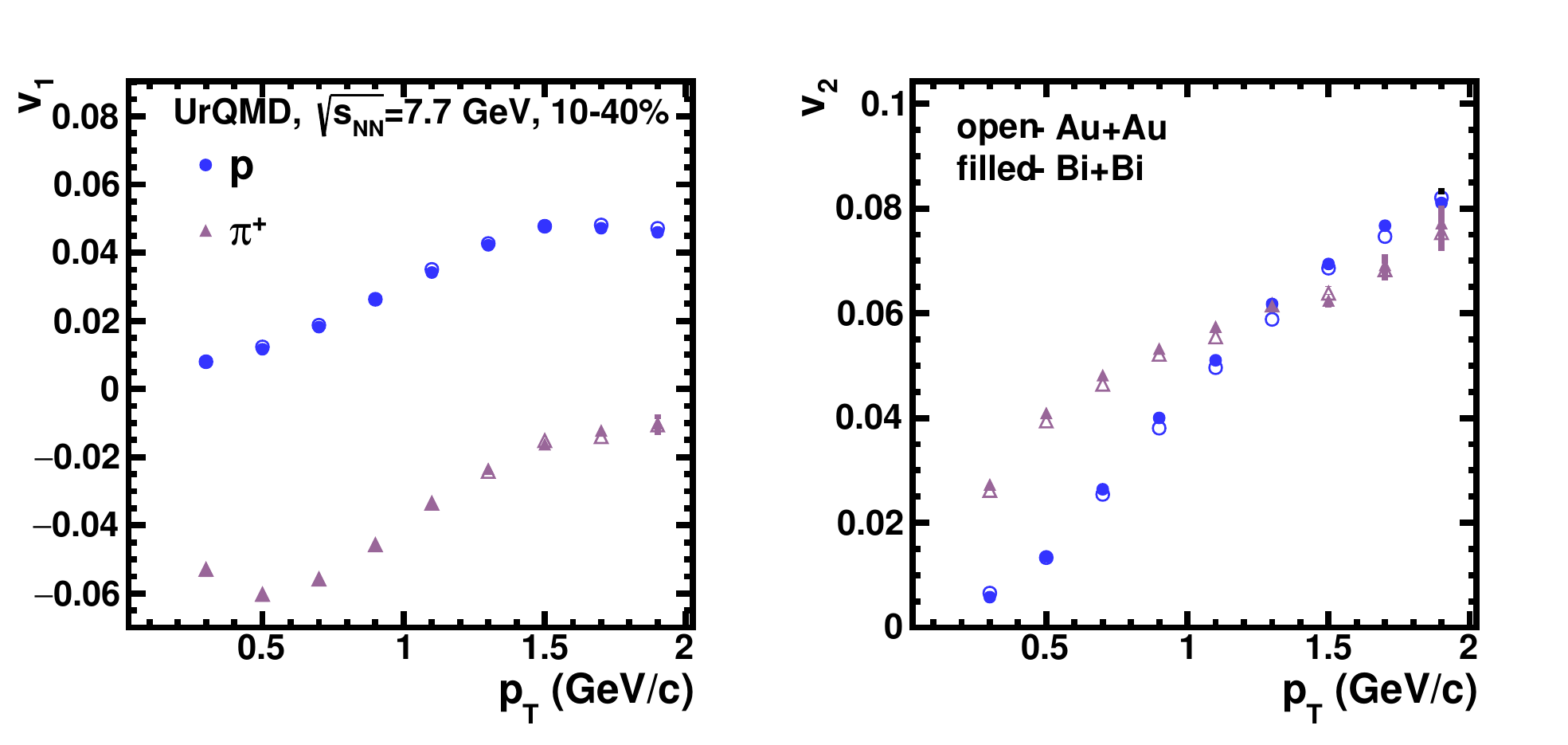}
	\vspace{-0.12in}
	\caption{$p_T$-dependence of directed $v_1$ (left) and elliptic $v_2$ (right) flow signals of pions and protons
		from 10-40\% midcentral Au+Au (open symbols) and Bi+Bi (filled symbols) collisions at $\sqrt{s_{NN}}=7.7$~GeV.}
	\label{fig:Sys_compare_pt}       
\end{figure}

Figure \ref{fig:Sys_compare_pt}  shows the MPD detector system's performance
for the  directed $v_1$ (left) and elliptic $v_2$ (right) flow
 measurements  of charged pions and protons from
10-40\% midcentral Au+Au (open symbols) and Bi+Bi (filled symbols) collisions at $\sqrt{s_{NN}}=7.7$~GeV.  The $v_n$ results
were obtained by event plane method: using the first order event plane ($\Psi_{1,\textrm{FHCal}}$) from FHCal.
In both cases, one can see the expected small difference between results for the event plane resolution and  $v_n$
between two colliding systems.

\section{Summary}
The MPD detector system's performance
for the diected ($v_1$) and elliptic  $v_2$ flow measurements of charged pions, protons, $K_{s}^{0}$ and $\Lambda$ particles
is studied with Monte-Carlo simulations using collisions of Au+Au and Bi+Bi  ions employing UrQMD  heavy-ion event generator.
We have shown how the various experimental measures
of elliptic flow are affected by fluctuations and nonflow at NICA energies. The detailed comparison of the $v_n$
results obtained from the analysis of the fully reconstructed data and
generator-level data allows to conclude that MPD system will allow reconstruction of
$v_n$ coefficients with high precision. 

\section{Acknowledgments}

This work is supported by the RFBR according to the research project No. 18-02-40086, the
European Union‘s Horizon 2020 research and innovation program under grant agreement No.
871072, by the Ministry of Science and Higher Education of the Russian Federation,
Project "Fundamental properties of elementary particles and cosmology" No 0723-2020-0041.

\end{document}